\documentclass[12pt]{article}
\usepackage{graphicx}
\usepackage{url}

\newcommand{\ttbar}{$t\bar{t}$ }
\newcommand{\stt}{$\sigma_{t\bar{t}}$ }
\newcommand{\fr}{\textit{right}}
\newcommand{\fl}{\textit{left}}

\newcommand\pubnumber{ATLAS and CMS}
\newcommand\pubdate{\today}

\def\institute{on behalf the ATLAS and CMS Collaborations\footnote{Copyright 2021 CERN for the benefit of the ATLAS and CMS Collaborations. Reproduction of this article or parts of it is allowed as specified in the CC-BY-4.0 license.}}
\def\support{\footnote{Università ``La Sapienza'' e INFN Roma 1}}

\def\Title#1{\begin{center} {\Large #1 } \end{center}}
\def\Author#1{\begin{center}{ \sc #1} \end{center}}
\def\Address#1{\begin{center}{ \it #1} \end{center}}

\newcommand\pubblock{\rightline{\begin{tabular}{l} \pubnumber\\
         \pubdate  \end{tabular}}}
\newenvironment{Abstract}{\begin{quotation}  }{\end{quotation}}
\newenvironment{Presented}{\begin{quotation} \begin{center} 
             PRESENTED AT\end{center}\bigskip 
      \begin{center}\begin{large}}{\end{large}\end{center} \end{quotation}}

\def\beq{\begin{equation}}
\def\eeq#1{\label{#1}\end{equation}}
\def\eeqn{\end{equation}}

\def\beqa{\begin{eqnarray}}
\def\eeqa#1{\label{#1}\end{eqnarray}}
\def\eeqan{\end{eqnarray}}

\let\bar=\overbar

\def\Dslash{\not{\hbox{\kern-4pt $D$}}}
\def\dslash{\not{\hbox{\kern-2pt $\del$}}}

\def\msb{{\bar{\ssstyle M \kern -1pt S}}}

\begin{document}
\begin{titlepage}
\pubblock

\vfill
\Title{Inclusive \& differential cross-section measurements of top-quark pair production with ATLAS and CMS}
\vfill
\Author{Luca Martinelli\support}
\Address{\institute}
\vfill
\begin{Abstract}
Latest results on inclusive top-quark pair production cross-sections are presented using collision data collected by ATLAS and CMS experiments at the LHC. Inclusive and differential measurements of top-quark pair production cross-sections from ATLAS and CMS are presented in the resolved and boosted kinematic regions. The cross-sections are measured as a function of various kinematic observables of the top quarks, the jets and leptons of the event final state.
\end{Abstract}
\vfill
\begin{Presented}
$14^\mathrm{th}$ International Workshop on Top Quark Physics\\
(videoconference), 13--17 September, 2021
\end{Presented}
\vfill
\end{titlepage}
\def\thefootnote{\fnsymbol{footnote}}
\setcounter{footnote}{0}

\section{Introduction}

The top-quark is the third-generation up-type quark and it is the heaviest known elementary particle. Good knowledge of the properties of the top-quark is essential because it introduces sizeable corrections in the electroweak parameters, with sizeable corrections.
A comprehensive set of measurements of the \ttbar production cross-sections ($\sigma_{t\bar{t}}$), both inclusive and differential, can be used to assess the current level of understanding of the Standard Model (SM), to perform studies to improve the Monte Carlo (MC) description of the \ttbar final state, thus reducing the systematic uncertainties of measurements, and also to set limits on the existence of new physics. ATLAS~\cite{ATLAS:2008xda} and CMS~\cite{CMS:2008xjf} experiments at LHC provided a large number of results during the last years at the $\sqrt{s} = 5.02,\ 13$ TeV in different regions of the phase-space, considering different channels and topology. In the following, the techniques used by ATLAS and CMS to measure \stt are presented with an overview of the recent results of the two experiments. 
%

\section{Inclusive $\sigma_{t\bar{t}}$ measurements at $\sqrt{s} = 5.02$ TeV}
\label{sec:5TeV}

A measurement of \stt at $\sqrt{s} = 5.02$ TeV was performed by both ATLAS~\cite{ATLAS:2021xhc} and CMS~\cite{CMS:2021jig}.

The ATLAS measurement is made using a sample of $257\ \textrm{pb}^{-1}$ in the dilepton channel, where the $W$ bosons from both top quarks decay leptonically. Events are selected requiring an opposite-charge pair of leptons ($ee$, $\mu\mu$ or $e\mu$) and one or two $b$-tagged jets. The \stt measurement is based on counting the number of $b$-tagged jets and it is extracted simultaneously with a parameter sensitive to the $b$-tagging efficiency to limit the related uncertainty (0.1\%). No requirements are applied on the number of jets, bringing to very small related uncertainties (0.03\%).
The cross-section is $\sigma_{t\bar{t}} = 66.0\pm4.5\textrm{(stat)}\pm1.6\textrm{(syst)}\pm1.2\textrm{(lumi)}\pm0.2\textrm{(beam)}\ \textrm{pb}$, where the first uncertain originates from statistics, the second from experimental systematics, the third from the measurement of the integrated luminosity, and the fourth from the beam energy uncertainty, in perfect agreement with the theoretical NNLO prediction $\sigma_{t\bar{t}} = 68.2\pm4.8(\textrm{PDF})^{+1.9}_{-2.3}(\alpha_S)\ \textrm{pb}$.

The CMS measurement is performed with a sample of $302\ \textrm{pb}^{-1}$ using, as ATLAS, the dilepton channel. This analysis is also count-based. It is combined with the single-lepton channel~\cite{CMS:2017zpm} measurement using the BLUE combination~\cite{Valassi:2013bga}. The cross-section is measured to be $\sigma_{t\bar{t}} = 63.0\pm4.1\textrm{(stat)}\pm3.0(\textrm{syst+lumi})\ \textrm{pb}$, where the first uncertain originates from statistics and the second from experimental and luminosity systematics, in perfect agreement with the theoretical NNLO prediction.

\section{Inclusive $\sigma_{t\bar{t}}$ measurements at $\sqrt{s} = 13$ TeV}
\label{sec:13TeV}

Using the same technique explained in Section~\ref{sec:5TeV}, ATLAS uses the dilepton channel to perform the measurement of \stt at the centre-of-mass energy of 13 TeV~\cite{ATLAS:2019hau}. The partial Run-II dataset of $36.1\ \textrm{fb}^{-1}$ collected during the 2015-2016 period is used. This measurement is limited by the uncertainty on the integrated luminosity (1.9\%) with a small contribution to the uncertainty originating from jets (0.03\%). An impact of 0.67\% and 0.52\% on the total uncertainty is coming from the \ttbar modelling uncertainties and the normalisation of the tW background, respectively. The cross-section is measured to be $\sigma_{t\bar{t}} = 826.4\pm3.6\textrm{(stat)}\pm11.5\textrm{(syst)}\pm15.7\textrm{(lumi)}\pm1.9\textrm{(beam)}\ \textrm{pb}$, where the first uncertain originates from statistics, the second from experimental systematics, the third from the measurement of the integrated luminosity, and the fourth from the beam energy uncertainty, and it is the most precise inclusive measurement of \stt at 13 TeV so far (2.40\% of relative uncertainty). It is in agreement with the theoretical NNLO prediction $\sigma_{t\bar{t}} = 832^{+20}_{-29}(\textrm{scale})\pm35(\textrm{PDF}+\alpha_S)\ \textrm{pb}$.

CMS has measured \stt in dilepton final states containing one $\tau$ lepton decaying hadronically and one electron or muon, using $35.9\ \textrm{fb}^{-1}$ of data collected at 13 TeV~\cite{CMS:2019snc}. This is the first \stt measurement at 13 TeV containing $\tau$ leptons. One electron or muon, together with at least 3 jets, are required in this analysis. At least one of the jets must be $b$-tagged and the other two must be tagged as an hadronic $\tau$ decay. The jet triplet invariant mass is used to construct the signal-background discriminator: $D_{jjb} = \sqrt{(m_W-m_{jj})^2+(m_t-m_{jjb})^2}$ where $m_W$ and $m_t$ are the W boson and top-quark mass respectively, $m_{jj}$ is the invariant mass of the two non-$b$-tagged jets and $m_{jjb}$ is the invariant mass of the jet triplet. The cross-section is measured to be $\sigma_{t\bar{t}} = 781\pm7\textrm{(stat)}\pm62\textrm{(syst)}\pm20\textrm{(lumi)}\ \textrm{pb}$, where the first uncertain originates from statistics, the second from experimental systematics, and the third from the measurement of the integrated luminosity, and it is dominated by the systematic uncertainties related to the $\tau$ identification (4.5\%) and mis-identification (2.3\%).

A new top-quark pair in the single-lepton channel (lepton+jets) production measurement has been performed by the ATLAS experiment using the full Run-II dataset ($139\ \textrm{fb}^{-1}$) at 13 TeV~\cite{ATLAS:2020aln}. Exactly one lepton ($e$ or $\mu$) is required with at least four jets. Three categories are defined according to the jet and $b$-jet multiplicity in the event: exactly one $b$-tagged jet plus at least three additional jets (signal region 1, SR1), exactly two $b$-tagged jets plus exactly two additional jets (SR2) and, exactly two $b$-tagged jets plus at least three additional jets (SR3). A profile likelihood approach is used to fit three distributions, one per each SR: the aplanarity for SR1, the minimum invariant mass of all lepton-jet combinations in SR2 and the average $\Delta R$ between the three constituent jets originating from the hadronically decaying top-quark in SR3. The cross-section is measured to be $\sigma_{t\bar{t}} = 830\pm0.4\textrm{(stat)}\pm36\textrm{(syst)}\pm14\textrm{(lumi)}\ \textrm{pb}$, where the first uncertain originates from statistics, the second from experimental systematics, and the third from the measurement of the integrated luminosity, which is systematically limited by the \ttbar modelling (2.9\%) and by the jet reconstruction uncertainties (2.6\%).

\section{Differential $\sigma_{t\bar{t}}$ measurements at $\sqrt{s} = 13$ TeV}

\begin{figure}[!h!tbp]
\centering
\includegraphics[width=.33\textwidth]{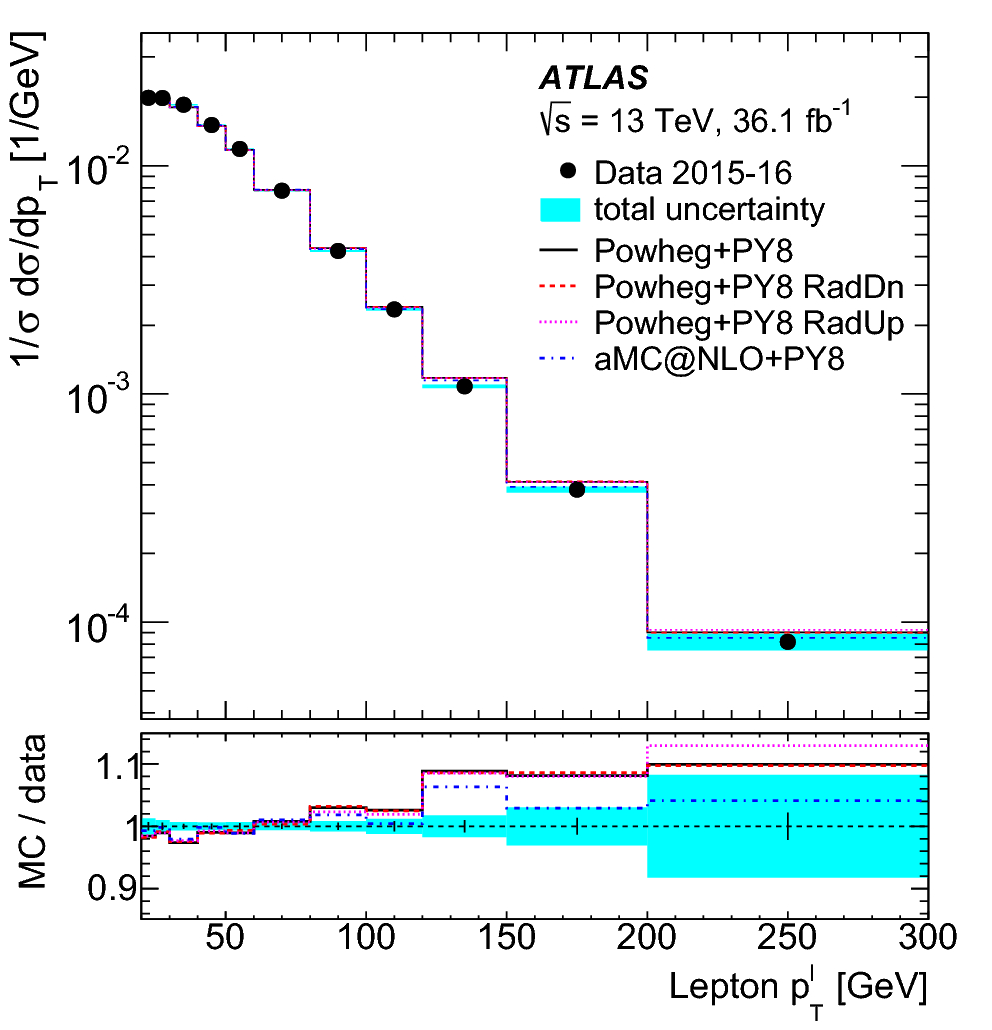} 
\qquad
\includegraphics[width=.58\textwidth]{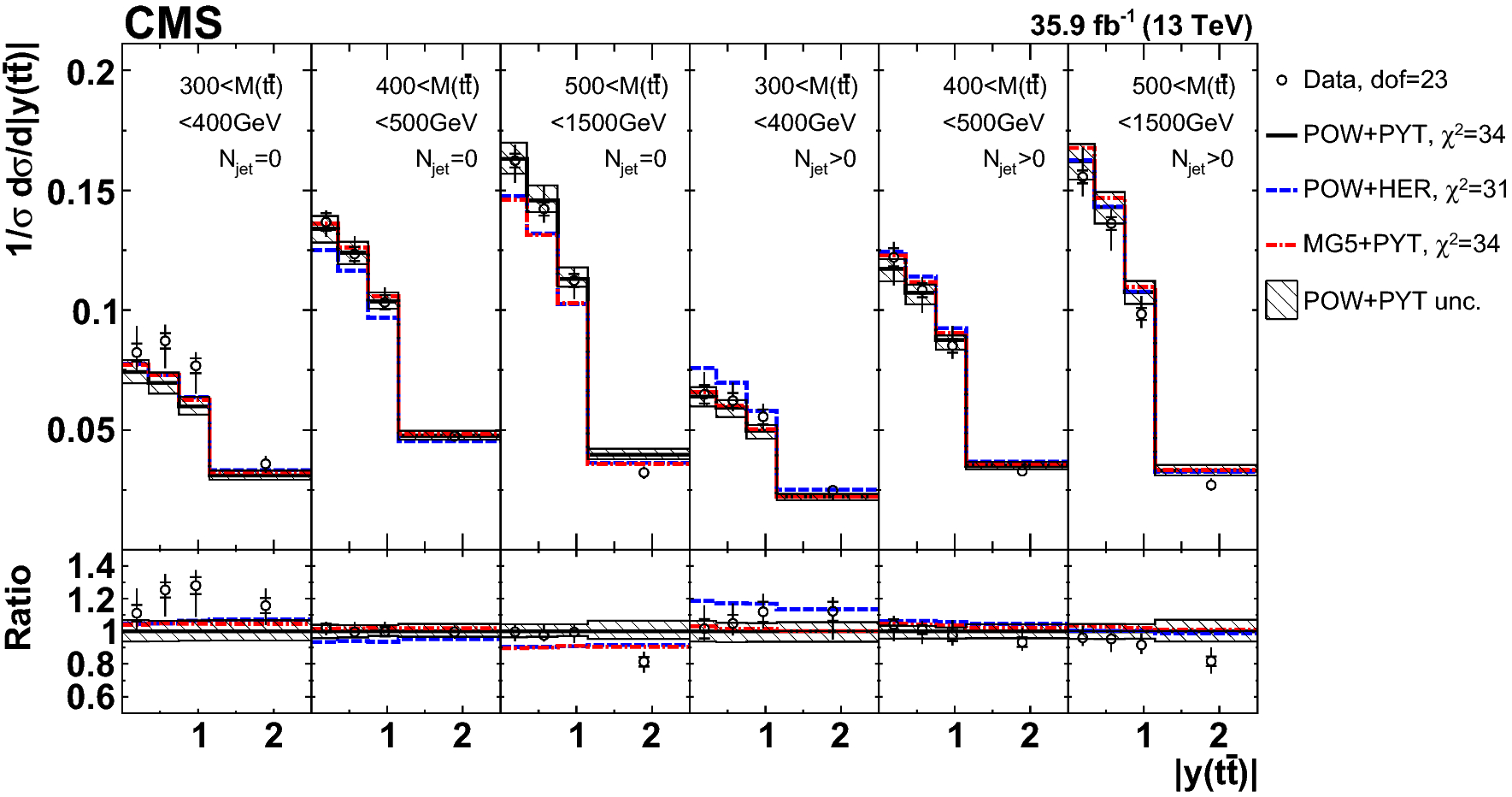}
\caption{\fl \ Normalised differential cross-section as a function of lepton $\textrm{p}_T$ measured by ATLAS~\cite{ATLAS:2019hau}. \fr \ Comparison of the measured number of jets, invariant mass of the \ttbar system and the rapidity of the \ttbar system cross-section to the theoretical predictions calculated using MC event generators, made by CMS~\cite{CMS:2019esx}.}
\label{fig:dilepton}
\end{figure}

The same analysis technique performed by ATLAS in the dilepton channel ($e\mu$), explained in Section~\ref{sec:13TeV}, is used to measure single- and double-differential \stt as a function of several lepton kinematics variables~\cite{ATLAS:2019hau}, as shown in Figure~\ref{fig:dilepton} (\fl). These measurements show that generators at NLO plus the parton shower (PS) give a good description of several observables but \textsc{Powheg} predicts a harder spectrum for the lepton $\textrm{p}_T$ variables.

In the \stt measurement in the dilepton channel performed by CMS~\cite{CMS:2019esx}, in addition to the $e\mu$, the $ee$ and $\mu\mu$ channels are considered. Double- and triple-differential measurements are performed at particle-level as shown in Figure~\ref{fig:dilepton} (\fr). The total uncertainties in these measurements are dominated by the systematic uncertainties and the largest one is related to the Jet energy scale. The best data-MC agreement is obtained by the \textsc{Powheg}+\textsc{Pythia} and \textsc{Powheg}+\textsc{Herwig} sample with \textsc{Powheg}+\textsc{Pythia} better in describing the measurements probing the jet multiplicity and extra radiation while \textsc{Powheg}+\textsc{Herwig} is better in the ones involving the $\textrm{p}_T$ distributions.

\begin{figure}[!h!tbp]
\centering
\includegraphics[width=.58\textwidth]{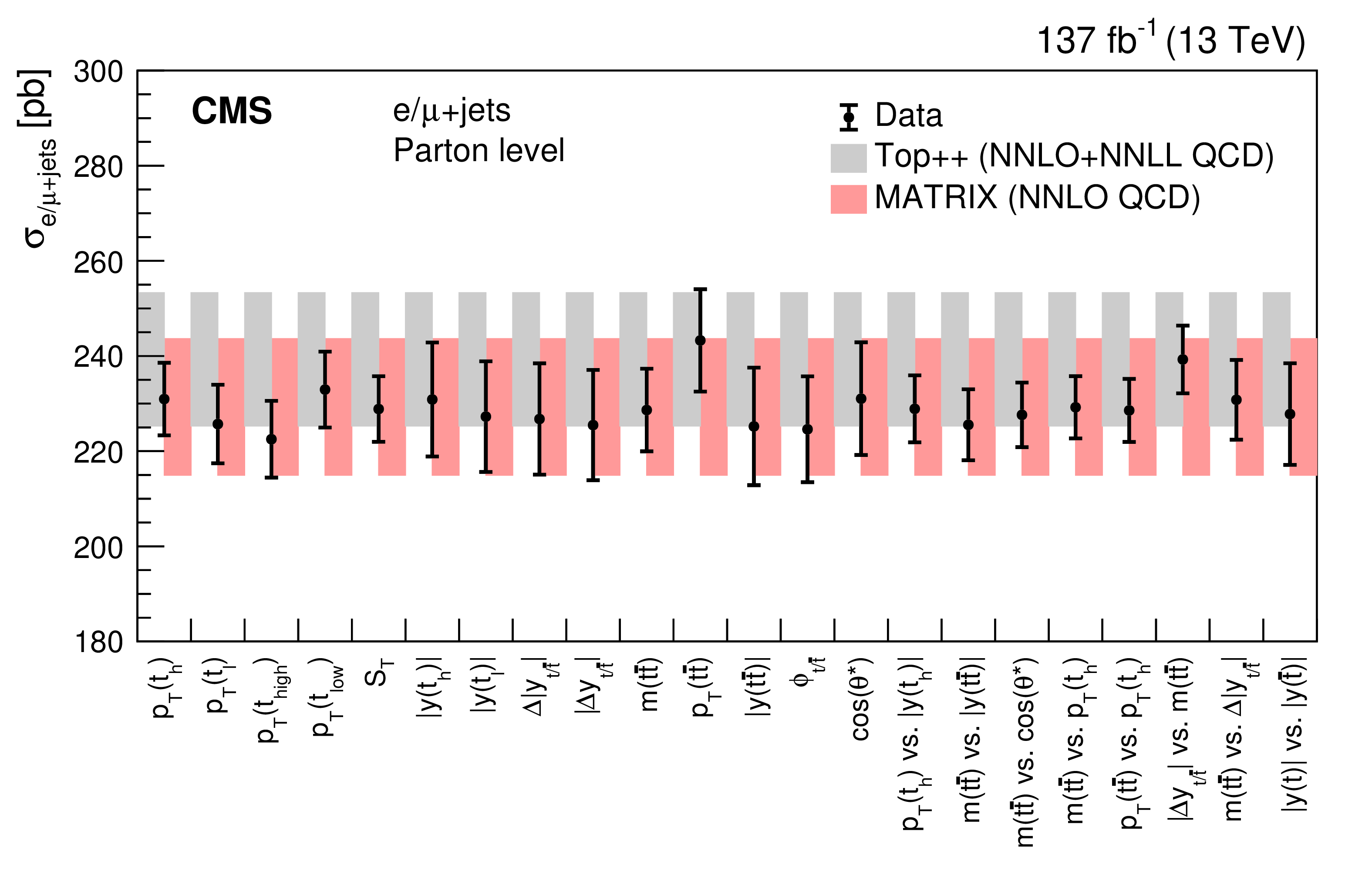} 
\caption{Measurements of the $\sigma^{t\bar{t}}_{e/\mu+\textrm{jets}}$ production cross-sections with their total uncertainty obtained as the sum of the cross-sections in all bins of a distribution as a function of the kinematic variable used in the determination~\cite{CMS:2021vhb}.} 
\label{fig:ljets_CMS}
\end{figure}

A new measurement with the full Run-II dataset has been performed by CMS in the single-lepton channel~\cite{CMS:2021vhb}. 
The two boosted categories are used for the background estimation while two resolved categories are used to reconstruct the system. More than twenty single- or double-differential measurements are performed with this technique, as shown in Figure~\ref{fig:ljets_CMS}. Summing all the bins, the inclusive \stt can be extracted: this is the best CMS measurement of \stt so far, with a total relative uncertainty of 3.2\%. The cross-section is measured to be $\sigma_{t\bar{t}} = 791\pm1\textrm{(stat)}\pm21\textrm{(syst)}\pm14\textrm{(lumi)}\ \textrm{pb}$, where the first uncertain originates from statistics, the second from experimental systematics, and the third from the measurement of the integrated luminosity. The data-MC agreement for the differential distributions is good for the \textsc{Powheg}+\textsc{Pythia} and \textsc{Powheg}+\textsc{Herwig} samples. A softer top $\textrm{p}_T$ is observed in data with respect to the MC predictions. The MC predictions fail to reproduce the jet observables.

\begin{figure}[!h!tbp]
\centering
\includegraphics[width=.31\textwidth]{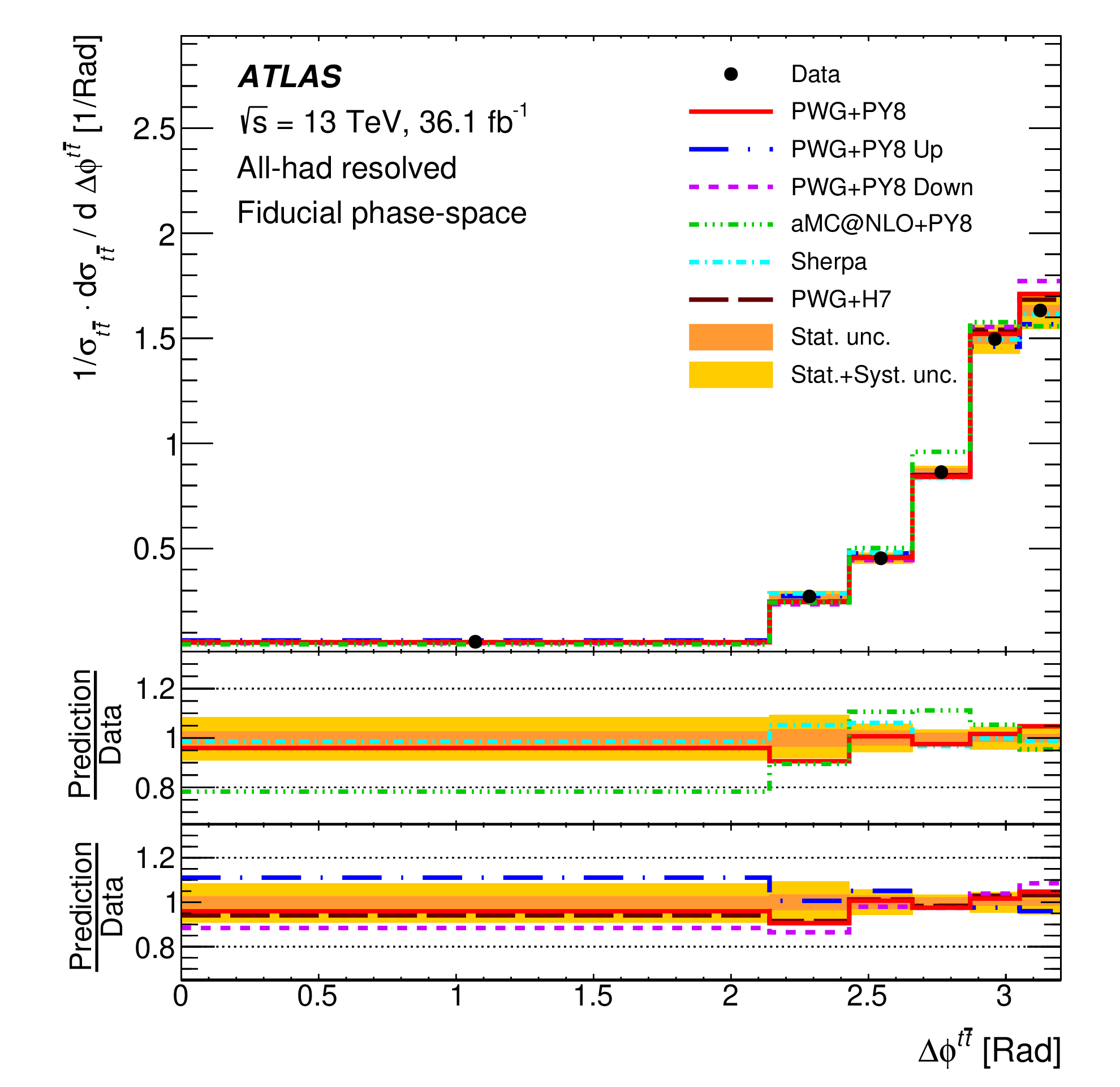} 
\includegraphics[width=.33\textwidth]{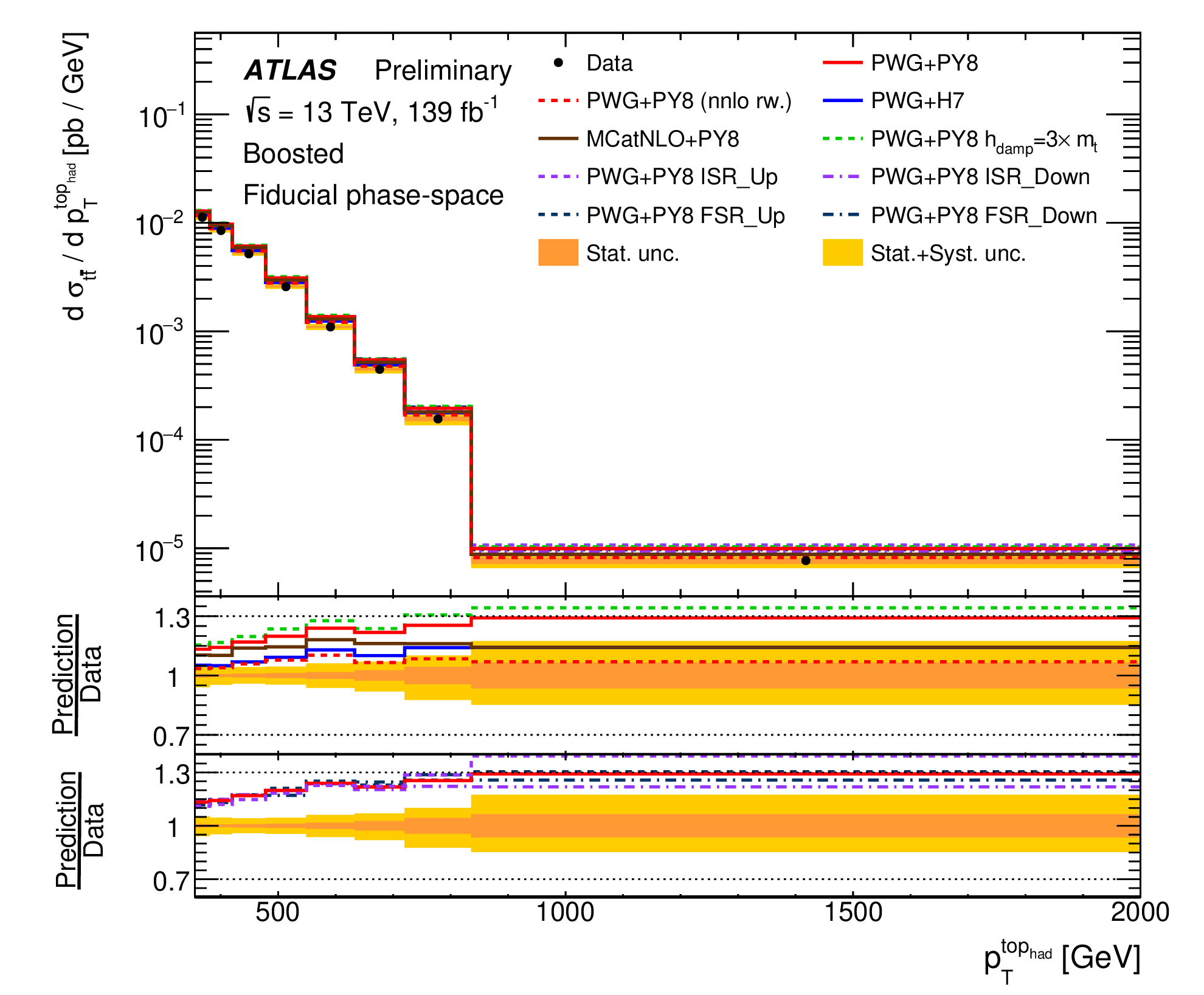} 
\includegraphics[width=.33\textwidth]{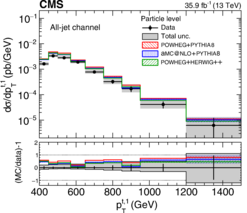} 
\caption{\fl \ Particle-level normalised single-differential cross-section as a function of the azimuthal separation $\Delta\Phi^{t\bar{t}}$ between the two top-quark candidates, measured by ATLAS~\cite{ATLAS:2020ccu}.\textit{(center)} Single-differential cross-section as a function of the hadronic top-quark $\textrm{p}_T$, measured by ATLAS~\cite{ATLAS:2021lxg}.\fr \ Differential cross-section unfolded to the particle-level as a function of the leading top-quark $\textrm{p}_T$, measured by CMS~\cite{CMS:2020tvq}.}
\label{fig:other_diff}
\end{figure}

The channel in which both top quarks decay hadronically is the most challenging given the abundant presence of hadrons in the event. The ATLAS measurement, shown in Figure~\ref{fig:other_diff} (\fl) , is performed requiring at least 6 jets in the event (with at least 2 $b$-jets) and making some kinematic requirements on system reconstruction, based on the reconstructed top-quark and W boson masses~\cite{ATLAS:2020ccu}. The dominant uncertainties on these measurements are from the PS and hadronisation modelling and from the multi-jet background estimate. The differential results show a good agreement with the NLO+PS generator on all the variables.

Thanks to the high energy provided by the LHC, a new kinematic region can be investigated, i.e. events with high-$\textrm{p}_T$ top quarks. In these cases, the decays of the top get a large boost and they are all contained in a cone with a radius of $\sqrt{\Delta\phi^2 + \Delta\eta^2} \sim$1. The large-R jets are used to reconstruct the hadronically decays of the top quarks. Both ATLAS~\cite{ATLAS:2021lxg} and CMS~\cite{CMS:2020tvq} perform measurements with this type of events, as shown in Figure~\ref{fig:other_diff} \textit{(center, right)}. The ATLAS measurement is performed in the single-lepton channel and it shows a top $\textrm{p}_T$ spectrum softer than MC predictions. If the MC predictions are re-weighted to the NNLO parton-level, the data-MC agreement improves. The differential measurements show also that the description of the additional radiation event is not good for all the tested generators but a good agreement in the shape of the distribution is visible. The CMS measurements are performed in the single-lepton channel and in the all-hadronic channel and they show that in both decay channels the observed absolute cross-sections are significantly lower than the predictions from theory, while the normalized differential measurements are well described.

\section{Conclusions}

Both ATLAS and CMS experiments at LHC have presented a significant number of inclusive and differential measurements of $\sigma_{t\bar{t}}$, with unprecedented precision reached in several channels. All the inclusive measurements are in agreement with the NNLL + NNLO calculations. Important differential measurements were made in all the channels both at particle- and parton-level. Future combinations of ATLAS and CMS differential results will represent a powerful instrument to verify the goodness of several MC generators and to improve the current knowledge on parton density functions inside protons.



\bibliographystyle{unsrt}
\typeout{}
\bibliography{eprint}

\end{document}